\documentclass[11pt]{article}
\usepackage{amsmath}
\usepackage{amsfonts}

\usepackage{url}

\newcommand{\bea}{\begin{eqnarray}}

\newcommand{\eea}{\end{eqnarray}}
\newcommand{\bean}{\begin{eqnarray*}}

\newcommand{\eean}{\end{eqnarray*}}
\newcommand{\be}{\begin{equation}}

\newcommand{\ee}{\end{equation}}

\begin{document}

\title{An example of an asymptotically $AdS_2\times S^2$ metric satisfying NEC but which is not exactly $AdS_2\times S^2$}
\author{Paul Tod\\
St John's College,\\St Giles',\\Oxford OX1 3JP}

\maketitle

\begin{abstract}
We give an example of an asymptotically $AdS_2\times S^2$ metric, in the sense of \cite{GG}, which satisfies the Null Energy Condition but is not exactly $AdS_2\times S^2$. It is therefore a counterexample to a conjecture of Maldacena mentioned in \cite{GG}, but it does not satisfy field equations. In an appendix we give an example admitting supercovariantly constant spinors as in \cite{t1}, which is asymptotically $AdS_2\times S^2$ on one side only.

\end{abstract}

\section{Introduction}
In a recent article, Galloway and Graf \cite{GG} have given a powerful structure theorem for metrics asymptotic to the standard metric on $AdS_2\times S^2$, also known as the Bertotti-Robinson metric, \cite{b,r}. They give their motivation as a desire to investigate a conjecture of Maldacena \cite{m} that such a metric, if it satisfies the Null Energy Condition (or \emph{NEC}, which we'll describe below), must be exactly $AdS_2\times S^2$. They are able to show that such a metric admits two foliations by shear-free, expansion-free null hypersurfaces, 
and any two of these, one from each foliation, intersect in a unit round $S^2$. Then they show that with the extra condition that the Ricci tensor is parallel (in the sense of covariantly constant) the metric becomes precisely $AdS_2\times S^2$. Without this extra condition, or for example some form of Einstein's equations with a source, it isn't clear from their work whether the NEC alone is enough too make the metric precisely $AdS_2\times S^2$. In this short note, we'll give an example to show that, in the absence of field equations, it is not.

For the example, we choose a simple metric ansatz which is compatible with the results of \cite{GG} and make an ad hoc assumption for the one free metric function to give the result. In an Appendix, we seek an example among the metrics found in \cite{t1}. While there isn't one we do find an interesting metric which is supersymmetric, in the sense of admitting supercovariantly constant spinors, globally nonsingular and asymptotically $AdS_2\times S^2$ on one side (in a sense to be made explicit).

\noindent{\bf{Acknowledgement:}} I am grateful to Greg Galloway and Melanie Graf for useful discussions and comments.

\section{The metric form considered}
Consider metrics of the form
\be\label{1}g=2A^2(u,v)dudv-4\frac{d\zeta d\overline\zeta}{P^2}\ee
with $P=1+\zeta\overline\zeta$. Then the second term is the unit 2-sphere metric in terms of the usual stereographic coordinate $\zeta=\tan(\theta/2)e^{i\phi}$, and the first term, with
\[u=(t+X)/\sqrt{2},\;\;\;v=(t-X)/\sqrt{2},\]
becomes
\be\label{1a}A^2(dt^2-dX^2).\ee
Put $x=\log\tan(X/2)$, so that $\cosh x=\mbox{csc} X$, to make this
\[\frac{A^2}{\cosh^2x}(\cosh^2xdt^2-dx^2),\]
so that (\ref{1}) is $AdS_2\times S^2$ or the Bertotti-Robinson metric \cite{b,r} when
\be\label{1aa}A^2=\cosh^2x=(\sin^2X)^{-1}.\ee
The space-coordinate ranges are $-\infty<x<\infty$ and correspondingly $0<X<\pi$ and clearly the space-time is foliated by two families of null hypersurfaces, of constant $u$ and constant $v$ respectively, meeting in unit two-spheres, as required by \cite{GG}. We don't yet know that the null hypersurfaces are shear-free and expansion-free, and there is no claim that this is the most general metric form consistent with their conditions, which would seem to be
\be\label{g1}
g=2\Theta dudv-4\left(\frac{d\zeta }{P}+Bdu+Cdv\right)\left(\frac{d\overline\zeta}{P}+\overline{B}du+\overline{C}dv\right),\ee
with the complex functions $B,C$ constrained further by the requirement that the null surfaces given by constant values of $u$ or $v$ are shear-free and expansion-free. We're restricting to the simpler case of $B=C=0$, $\Theta>0$ and spherical symmetry. Also we shall need asymptotic conditions on $A(u,v)$ at some stage.

\medskip

We want the connection coefficients and curvature components and a quick way to these is to use the NP formalism on (\ref{1}) using the null tetrad of forms
\[\ell=Adv,\;\;n=Adu,\;\;m=-\frac{\sqrt{2}d\overline\zeta}{P},\;\;\overline{m}=-\frac{\sqrt{2}d\zeta}{P}\]
with corresponding directional derivatives (i.e. the dual basis)
\[D=A^{-1}\partial_u,\;\;\Delta=A^{-1}\partial_v,\;\;\delta=\frac{P}{\sqrt{2}}\partial_\zeta,\;\;\overline\delta=\frac{P}{\sqrt{2}}\partial_{\overline\zeta},\]
and calculate the spin-coefficients to find
\[\kappa=\sigma=\rho=\pi=\nu=\lambda=\mu=\tau=0\]
which means that $\ell$ and $n$ are both geodesic, shear-free and expansion-free, as  desired, though they are not affinely parametrised. For the rest find
\[\epsilon=\frac{DA}{2A},\;\;\gamma=-\frac{\Delta A}{2A},\;\;\alpha=-\overline{\beta}=\frac{\zeta}{2\sqrt{2}}.\]
We proceed to the curvature components to find that the only nonzero ones can be given in terms of a single real function $Q$ as
\begin{eqnarray}\label{a1}
\Lambda&=& \frac{1}{12}+\frac16Q\\
\psi_2&=& -\frac16-\frac13Q \label{a2} \\
\phi_{11}&=&\frac14-\frac12Q\label{a3}
\end{eqnarray}
where
\be\label{2}Q=\frac{A_{uv}}{A^3}-\frac{A_uA_v}{A^4},\ee
which is one quarter of the scalar curvature of the metric in (\ref{1a}) (in conventions used here). Recall here the 4-dimensional Ricci scalar is $R=24\Lambda$, $\psi_2$ is the remaining nonzero component of the Weyl spinor corresponding to the Weyl tensor (which is therefore type D), and $\phi_{11}$ is the remaining nonzero component of the Ricci spinor, corresponding to the trace-free part of the Ricci tensor.

\medskip

It is not hard to see that the Ricci tensor is parallel just when $\phi_{11}$ is constant, which then also forces $Q,\Lambda$ and $\psi_2$ to be constant. This metric is then asymptotically-$AdS_2\times S^2$ only if $Q$ takes right value (of $-1/2$, given in the next subsection), when the metric is exactly $AdS_2\times S^2$, as it must be by \cite{GG}.
\subsection{The Bertotti-Robinson metric}
This is the Einstein-Maxwell solution in this family. For Einstein-Maxwell solutions $\Lambda$ is zero and then from the array (\ref{a1})-(\ref{a3}) therefore also $\psi_2=0$ (and the Bertotti-Robinson metric is conformally-flat, as is familiar; see e.g.  \cite{cf}). This forces $Q=-1/2$ and then $\phi_{11}=1/2$ which can be achieved with a constant magnetic field in the $x$-direction.

\medskip

In general the NEC is the statement
\[\phi_{ABA'B'}\alpha^A\alpha^B\overline{\alpha}^{A'}\overline{\alpha}^{B'}\geq 0\]
when $\phi_{ABA'B'}$ is the Ricci spinor and $\alpha^A$ is an arbitrary spinor, but here the only nonzero component of the Ricci spinor is $\phi_{11}$ so that, in the spinor dyad implied by the NP tetrad we're using:
\[\phi_{ABA'B'}=4\phi_{11}o_{(A}\iota_{B)}\overline{o}_{(A'}\overline{\iota}_{B')},\]
and then, decomposing $\alpha$ in the dyad as $\alpha^A=Xo^A+Y\iota^A$ we find
\[\phi_{ABA'B'}\alpha^A\alpha^B\overline{\alpha}^{A'}\overline{\alpha}^{B'}=4\phi_{11}|X|^2|Y|^2,\]
which is nonnegative provide $\phi_{11}$ is. Thus the NEC here is just the requirement
\be\label{nec}\phi_{11}\geq 0,\ee
or equivalently $Q\leq 1/2$.

\medskip

Many solutions for $A(u,v)$ are possible but a simple one is to make it independent of $t$, say $A=e^{\phi(X)}$. With
\[\partial_u=\frac{1}{\sqrt{2}}(\partial_t+\partial_X),\;\;\;\partial_v=\frac{1}{\sqrt{2}}(\partial_t-\partial_X)\]
we get
\[Q=-\frac12e^{-2\phi}\phi_{XX}=-\frac12,\]
for Bertotti-Robinson, when
\be\label{3}\phi=-\log\sin X,\ee
will do, and this agrees with the metric (\ref{1aa}).

\medskip

Now we  want something close to this but still satisfying the NEC.
\subsection{Perturbed Bertotti-Robinson metric}
We try
\be\label{4}
\phi(X)=-\log\sin X -f(X), \ee
when NEC (\ref{nec}) requires
\[e^{2f}(1-\sin^2X f'')\geq -1.\]
Again a simple choice is
\[f(X)=\epsilon X(\pi-X),\]
chosen to vanish at the two boundaries $X=0,\pi$. Then NEC requires
\[ e^{2\epsilon X(\pi-X)}(1+2\epsilon\sin^2X)\geq -1,\]
which is clearly satisfied for $0<X<\pi$ with positive $\epsilon$.

Does this choice satisfy the asymptotic fall-off conditions (2.1)-(2.4) of \cite{GG}?
\subsection{Checking the asymptotic conditions of \cite{GG}}
In the terminology of \cite{GG} $\mathring{g}$ is the $AdS_2\times S^2$ metric, and $h$ is the difference between this and the metric considered:
\[h=g-\mathring{g}=\frac{1}{\sin^2X}(e^{2f}-1)(dt^2-dX^2). \]
 One chooses an orthonormal basis for $\mathring{g}$, which can be $(e_0,e_1)=(\sin X\partial_t,\sin X\partial_X)$ (we don't need to specify $e_2$ or $e_3$ as $h$ is orthogonal to both and independent of $\zeta$). Condition ($b_1$) of \cite{GG} invites us to consider
\[h_{00}=h(e_0,e_0)=-h_{11}=-h(e_1,e_1)=e^{2f}-1,\mbox{   while  }h_{01}=0.\]
In equation (2.1) of \cite{GG} we need the function $|x|$ and recall from above that this is $|\log\tan(X/2)|$, so to satisfy (2.1) of \cite{GG} we need a (real, positive) constant $c$ such that
\be\label{b1} |\log\tan(X/2)|(e^{2f}-1)\leq c,\ee
and this can be achieved as the product on the LHS is bounded on the approach to $X=0$ or $\pi$ and smooth elsewhere. Next
for (2.2) in \cite{GG} the $e_0$-derivative is zero while the $e_1$-derivative requires a constant $C_1$ with
\be\label{b2} \sin X|\log\tan(X/2)||2f_X|e^{2f}\leq C_1,\ee
and the troublesome term $\sin X|\log\tan(X/2)|$ is again bounded at $X=0,\pi$ and smooth elsewhere. Finally for (2.3) in \cite{GG}, again $e_0$ derivatives are trivial, and the only requirement is that $C_1$ also satisfies
\be\label{b3} 2e^{2f}\sin X|\log\tan(X/2)||\sin X(f_{XX}+2f_X^2)+\cos Xf_X|\leq C_1.\ee
This has the same possibly troublesome term $\sin X|\log\tan(X/2)|$, but which is bounded at $X=0,\pi$.

\medskip

We've checked that this metric is asymptotically $AdS_2\times S^2$ but it isn't exactly $AdS_2\times S^2$ as the curvature is different -- in particular it isn't conformally-flat.

\section*{Appendix: the metrics of \cite{t1}}
$AdS_2\times S^2$, as the Bertotti-Robinson solution, lies in the family of (four-dimensional) metrics admitting supercovariantly constant spinors, which can be called supersymmetric for this reason, that were considered in \cite{t1}. One might wonder whether any others of this class might provide counter-examples to the conjecture considered here.

Recall in the static case from \cite{t1} the metric takes the form
\[g=V^2dt^2-V^{-2}(dx^2+dy^2+dz^2),\]
with $V(x,y,z)$ and
\[\nabla^2(V^{-1})=-\rho V^{-1},\]
for a real non-negative function $\rho$, where $\nabla^2$ is the flat Laplacian in $(x,y,z)$. There is a Maxwell field
\[F=c_1dV\wedge dt,\]
or the dual of this (since duality leaves the energy-momentum tensor unchanged) where $c_1$ is a convention-dependent constant.

This metric satisfies the Einstein equations with source a sum of two terms: a charged dust with 4-velocity $V^{-1}\partial/\partial t$ and both mass and charge density equal to $\rho V^2$; and the energy-momentum tensor of the electromagnetic field above, which has the current of the charged dust as its source. The form of the Einstein tensor guarantees that the NEC is satisfied provided $\rho\geq 0$.

\medskip

In 3-dimensional spherical polars and with $V(r)$ the metric becomes
\[g=V^2dt^2-V^{-2}(dr^2+r^2(d\theta^2+\sin^2\theta d\phi^2)),\]
when the choice $V=r$ has vanishing $\rho$ and in fact gives the Bertotti-Robinson metric. By duality one can regard the Maxwell field as purely magnetic, which was the interpretation given by Bertotti \cite{b} and Robinson \cite{r}. It's hard to see another choice of $V$ giving the metric (\ref{1}) with a different choice of $A$ but there is an interesting choice that leaves the $AdS_2$-part of the metric unchanged, namely
\[V=(r^2+a^2)^{1/2}\mbox{   when  }\rho=\frac{3a^2}{(r^2+a^2)^2}.\]
The metric is now
\be\label{A1}g=(r^2+a^2)dt^2-\frac{1}{r^2+a^2}(dx^2+dy^2+dz^2),\ee
which is defined on $\mathbb{R}^4$ and in spherical polars is
\be\label{A2}g=(r^2+a^2)dt^2-\frac{dr^2}{r^2+a^2}-\frac{r^2}{r^2+a^2}(d\theta^2+\sin^2\theta d\phi^2).\ee
From the form (\ref{A1}) we know that the coordinate singularity at $r=0$ is removable.

The metric (\ref{A2}) is in fact conformally related to the Bertotti-Robinson metric, by conformal factor $\Omega^2=(r^2+a^2)/r^2$ accompanied by a rescaling of $t$, and so it is again conformally flat.

Set $r=a\sinh x, t=\tau/a$ to convert (\ref{A2}) to
\be\label{A3}g=\cosh^2xd\tau^2-dx^2-\tanh^2x(d\theta^2+\sin^2\theta d\phi^2),\ee
which is a warped product of $AdS_2$ and $S^2$. As we've seen, the apparent metric singularity at $x=0$ is just a coordinate singularity and corresponds to the origin of spatial coordinates. Now the metric (\ref{A2}) does not take the form found by Galloway and Graf (\ref{g1}) and is not in fact asymptotically $AdS_2\times S^2$ as previously defined: it has one end at $x\rightarrow\infty$ where the asymptotic conditions in \cite{GG} hold, but the other end as $x\rightarrow -\infty$ is missing, cut off by the origin of $x$.

\end{document}